\documentclass[journal]{IEEEtran}

\usepackage{amsmath}
\usepackage{graphicx}
\usepackage{url}
\usepackage{hyperref}
\usepackage{cite}
\usepackage{amsfonts}
\usepackage{mathtools}
\usepackage{hhline}
\usepackage[caption=false, font=footnotesize]{subfig}
\usepackage{tabstackengine}
\usepackage{algpseudocode}
\usepackage{algorithm}
\algnewcommand\algorithmicforeach{\textbf{for each}}
\algdef{S}[FOR]{ForEach}[1]{\algorithmicforeach\ #1\ \algorithmicdo}

\usepackage[table]{xcolor}
\usepackage{commath,multirow}
\usepackage{tabularx} 
\usepackage{diagbox}

\usepackage{verbatim} 

\usepackage{adjustbox,cleveref}

\newcommand{\ie}{{i}.{e}.}

\makeatletter
\long\def\@makecaption#1#2{\ifx\@captype\@IEEEtablestring%
\footnotesize\begin{center}{\normalfont\footnotesize #1}\\
{\normalfont\footnotesize\scshape #2}\end{center}%
\@IEEEtablecaptionsepspace
\else
\@IEEEfigurecaptionsepspace
\setbox\@tempboxa\hbox{\normalfont\footnotesize {#1.}~~ #2}%
\ifdim \wd\@tempboxa >\hsize%
\setbox\@tempboxa\hbox{\normalfont\footnotesize {#1.}~~ }%
\parbox[t]{\hsize}{\normalfont\footnotesize \noindent\unhbox\@tempboxa#2}%
\else
\hbox to\hsize{\normalfont\footnotesize\hfil\box\@tempboxa\hfil}\fi\fi}
\makeatother

\begin{document}
%
\title{Making Video Quality Assessment Models Sensitive to Frame Rate Distortions}

\author{Pavan C. Madhusudana, Neil Birkbeck, Yilin Wang,  Balu Adsumilli and Alan C. Bovik 
	\thanks{P. C. Madhusudana and A. C. Bovik are with the Department of Electrical and
Computer Engineering, University of Texas at Austin, Austin, TX, USA (e-mail:
pavancm@utexas.edu; bovik@ece.utexas.edu). Neil Birkbeck, Yilin Wang
and Balu Adsumilli are with Google Inc. (e-mail: birkbeck@google.com; yilin@google.com; badsumilli@google.com).}}


\maketitle


\begin{abstract}
We consider the problem of capturing distortions arising from changes in frame rate as part of Video Quality Assessment (VQA). Variable frame rate (VFR) videos have become much more common, and streamed videos commonly range from 30 frames per second (fps) up to 120 fps. VFR-VQA offers unique challenges in terms of distortion types as well as in making non-uniform comparisons of reference and distorted videos having different frame rates. The majority of current VQA models require compared videos to be of the same frame rate, but are unable to adequately account for frame rate artifacts. The recently proposed Generalized Entropic Difference (GREED) VQA model succeeds at this task, using natural video statistics models of entropic differences of temporal band-pass coefficients, delivering superior performance on predicting video quality changes arising from frame rate distortions. Here we propose a simple fusion framework, whereby temporal features from GREED are combined with existing VQA models, towards improving model sensitivity towards frame rate distortions. We find through extensive experiments that this feature fusion significantly boosts model performance on both HFR/VFR datasets as well as fixed frame rate (FFR) VQA databases. Our results suggest that employing efficient temporal representations can result much more robust and accurate VQA models when frame rate variations can occur.
\end{abstract}


\section{Introduction}
\IEEEPARstart{T}{he} increased popularity of streaming, entertainment and social media content has garnered significant interest towards the development of methods that can improve perceptual video quality and provide better consumer experiences. Some of the approaches that have been explored in the past include higher spatial resolutions (4K/8K) \cite{ge2017toward}, high dynamic range (HDR) \cite{mai2010optimizing}, immersive content (virtual reality/augmented reality) \cite{smolic2007coding} and increasing temporal resolutions by employing high frame rate (HFR) videos. Among these, improving quality via HFR has been relatively less explored, as the majority of the currently available content is limited to 60 frames-per-second (fps). This has begun to change however, given the availability of high-speed cameras and displays, and the desire to stream high motion content, such as live sports.

A variety of factors have contributed to the limited adoption of HFR videos. In the past, capturing HFR content required expensive cameras which were not designed for television formats, thereby reducing the availability of HFR sequences. Further, display devices capable of supporting HFR content were less commonly available. Further, the perceptual benefits associated with higher frame rate were not clearly understood. However, recently advancements in both capture and display technologies have made HFR videos much more viable, thus increasing their potential use.

As the video industry transitions towards HFR regimes, it is important to systematically investigate and analyze the advantages associated with HFR videos. Further, perceptually optimizing modern bit rate ladders to accommodate frame rate changes, much as spatial downscaling is used now, could significantly improve streaming efficiency. Of course, adopting HFR implies significant potential increases in bandwidth consumption, hence is necessary to understand the merits of HFR. A common notion is that HFR videos tend to have better quality as the number of frames is increased, since it can often replicate motion more smoothly. However, very few studies have systematically evaluated these notions. The databases which are directed towards understanding and modeling the perceptual implications of frame rate include Waterloo HFR \cite{nasiri2015perceptual}, BVI-HFR \cite{mackin2018study} and LIVE-YT-HFR \cite{madhusudana2021liveythfr}, all of which contain Variable Frame Rate (VFR) and/or HFR videos.

Objective Video Quality Assessment (VQA) models are designed to predict human perception of video quality. VQA models can be broadly categorized into three types : Full Reference (FR) models which have access to pristine high quality reference signals \cite{wang2004image,wang2003multiscale,VMAF2016}, Reduced Reference (RR) models which use partial reference information \cite{soundararajan2012rred,soundararajan2012video}, and No Reference (NR) models which only have the distorted videos to make quality predictions. VFR-VQA problem is concerned with comparing reference and distorted videos at possibly difference frame rates, thus can be categorized as either FR or RR VQA models.

Existing FR VQA models typically require the pristine and distorted sequences being compared to share same frame rate, thus limiting their direct applicability to VFR-VQA problem. Although these methods can be extended to VFR videos by employing suitable preprocessing techniques, such as temporal upsampling of the distorted sequence, it has been observed that these operations often introduced additional artifacts, leading to inferior correlations against human judgments \cite{madhusudana2021liveythfr}. Moreover, these FR models are mainly intended to account for commonly observed artifacts such as compression, blur, scaling etc. but not those arising due to changes in frame rate.

Frame rate distortions are spatio-temporal, but the temporal aspects are more evident since these artifacts arise due to temporal subsampling. Thus, it is imperative to formulate effective, perceptually relevant temporal representations to achieve better quality predictions on VFR videos. Existing methods generally place less emphasis on temporal aspects of visual quality, usually employing frame differences \cite{soundararajan2012video,bampis2017speed,VMAF2016,bampis2018spatiotemporal}, which are generally insufficient to capture the complex variations, often spanning multiple frame instants, that can arise due to frame rate changes. Towards advancing progress on these problems, here we describe and implement an effective way to incorporate temporal features into existing VQA models, making them more sensitive to frame rate induced artifacts.

Our proposed model uses a fusion framework, whereby we combine spatial features from any existing FR-VQA model with temporal features from the recently developed GREED \cite{madhusudana2021greed} model. The GREED model has been shown to achieve superior correlations against subjective judgments of VFR video quality, when evaluated on videos of varied frame rates. In \cite{madhusudana2021pcs}, a preliminary version of this current work, we experimented with fusing GREED and Video Multi-method Fusion (VMAF) \cite{VMAF2016} features towards capturing frame rate dependent artifacts. The results we achieved were provocative enough to motivate the current greatly expanded study. Here, we extend this idea and show that this type of feature fusion can be generalized to any existing FR VQA models, including FR Image Quality Assessment (IQA) models resulting in significant performance gains in capturing the perceptual effects of frame rate distortions. The design framework is quite simple, it effectively promotes the advantages of both the fused models, resulting in robust performance on both HFR/VFR and non-VFR VQA standard databases.


\begin{algorithm}[t] \caption{GST Feature Fusion}
\begin{algorithmic}[1]
\Require reference video $R$, distorted video $D$, VQA model F
\Ensure GST-F features
\State Calculate spatial index from model F, $\text{S} = \text{F}(R,D)$
\State Temporally subsample $R$ to get $PR$
\State Scale $S = \{4,5\}$, band-pass filter bank $b_k: k \in \{1,\ldots 7\}$
\ForEach {$s \in S $}
\ForEach {$b_k$}
\State Calculate TGREED$_k$ from (\ref{eqn:GTI})
\EndFor
\EndFor
\State Concatenate S and TGREED to obtain GST-F features.
\end{algorithmic}
\label{alg1}
\end{algorithm}

\section{Background and Method}
\label{sec:GREED}
Our proposed method involves employing the spatial components from existing FR VQA models, and the temporal component of GREED. In our experiments we use four leading VQA models : SSIM \cite{wang2004image}, MS-SSIM \cite{wang2003multiscale}, ST-RRED \cite{soundararajan2012video} and SpEED \cite{bampis2017speed}. The framework is fairly generic in nature and can easily be extended to other FR VQA models. In this section we first provide a brief overview of temporal GREED, followed by a discussion of our proposed feature fusion methodology.

\begin{table}[t]
\caption{Performance comparison of FR algorithms on the LIVE-YT-HFR Database. The reported numbers are median values from 200 iterations of randomly chosen train-test sets. In each column the first and second best models are boldfaced. We obtained numbers from \cite{madhusudana2021greed,madhusudana2021pcs} for the results of GREED and GREED-VMAF.}
    \label{Table:MOS_comparison}
    \centering
    \begin{tabular}{|c||c|c|c|c|}
        \hline
        ~    & SROCC $\uparrow$ & KROCC $\uparrow$ & PLCC $\uparrow$ & RMSE $\downarrow$ \\ \hline \hline
        SSIM \cite{wang2004image} & 0.5566 & 0.4042 & 0.5418 & 9.99 \\
        GST SSIM & 0.7576 & 0.5648 & 0.7700 & 7.40 \\ \hline
        MS-SSIM \cite{wang2003multiscale} & 0.5742 & 0.4135 & 0.5512 & 10.01 \\
        GST MS-SSIM & 0.8128 & 0.6204 & 0.8179 & 6.75 \\ \hline
        ST-RRED \cite{soundararajan2012video} & 0.6394 & 0.4516 & 0.6073 & 9.58 \\
        GST ST-RRED & 0.8029 & 0.5937 & 0.8144 & 6.91 \\ \hline
        SpEED \cite{bampis2017speed} & 0.6051 & 0.4437 & 0.5206 & 10.28 \\
        GST SpEED & 0.8276 & 0.6343 & 0.8346 & 6.51 \\ \hline
        VMAF \cite{VMAF2016}& 0.7782 & 0.5918 & 0.7419 & 8.10 \\
        GREED & \textbf{0.8822} & \textbf{0.7046} & \textbf{0.8869} & \textbf{5.48} \\
        GREED-VMAF & \textbf{0.8658} & \textbf{0.6840} & \textbf{0.8723} & \textbf{5.89} \\
        \hline
    \end{tabular}
\end{table}

\vspace{-7pt}
\subsection{Temporal GREED (TGREED)}
GREED model is motivated by deviations that occur in the statistical distributions of temporal band-pass video coefficients that arise as frame rates are varied. In particular TGREED, captures the temporal aspects of frame rate distortions by employing a bank of temporal band-pass filters. The video $V(\mathbf{x},t)$ ($\mathbf{x} = (x,y)$ is subjected to temporal band-pass decomposition by a bank of $K$ 1D filters denoted by $b_k$ for $k \in \{1,\ldots K\}$
\begin{align}
    B_k(\mathbf{x},t) = V(\mathbf{x},t)*b_k(t) \text{\hspace{10pt}} \forall k \in \{1,\ldots K\},
\end{align}
where $B_k$ denotes the band-pass response of a video to the $k^{th}$ filter. The distributions of $B_k$ reliably exhibit heavy tailed shapes and can be modeled as following a Generalized Gaussian Distribution (GGD). Deviations from these model distributions caused by distortion are quantified by comparing the entropies of the reference and distorted sequences. The entropy of the GGD is given by 
\begin{align}
    h(X) = \frac{1}{\beta} - \log \left(\frac{\beta}{2\alpha\Gamma(1/\beta) }\right).
    \label{eqn:ggd_entropy}
\end{align}
where $\beta$ and $\alpha$ are parameters of the GGD, and $\Gamma$ denotes the Gamma function:
\begin{align*}
    \Gamma(a) = \int_0 ^{\infty}x^{a-1}e^{-x} dx .
\end{align*}
The band-pass coefficients $B_k$ are partitioned into non-overlapping patches of size $\sqrt{M} \times \sqrt{M}$, which are indexed by $p \in \{ 1,2, \ldots P \}$. In our experiments we employed spatial patches of size $5 \times 5$ (\ie{ }$\sqrt{M} = 5$). The GGD parameters $\beta$ and $\alpha$ are computed on each patch using the kurtosis matching procedure detailed in \cite{soury2015new,pan2012exposing}.

\begin{table*}[t]
	\caption{Performance comparison of FR methods for specific frame rates on the LIVE-YT-HFR Database. The reported numbers are median values over 200 iterations of randomly chosen train-test sets. In each column the first and second best values are marked in boldface. We obtained numbers from \cite{madhusudana2021greed,madhusudana2021pcs} for the results of GREED and GREED-VMAF.}
	\label{Table:FPS_comparison}
	\centering
	\scriptsize
	\scalebox{1.02}{
		\begin{tabular}{|c||c|c|c|c|c|c|c|c|c|c|c|c|}
			\hline
			& \multicolumn{2}{|c|}{24 fps} & \multicolumn{2}{|c|}{30 fps} & \multicolumn{2}{|c|}{60 fps} & \multicolumn{2}{|c|}{82 fps} & \multicolumn{2}{|c|}{98 fps} & \multicolumn{2}{|c|}{120 fps} \\
			\cline{2-13}
			~ & SROCC$\uparrow$ & PLCC$\uparrow$ & SROCC$\uparrow$ & PLCC$\uparrow$ & SROCC$\uparrow$ & PLCC$\uparrow$ & SROCC$\uparrow$ & PLCC$\uparrow$ & SROCC$\uparrow$ & PLCC$\uparrow$ & SROCC$\uparrow$ & PLCC$\uparrow$ \\ \hline \hline
			SSIM \cite{wang2004image} & 0.266 & 0.222 & 0.283 & 0.189 & 0.382 & 0.302 & 0.371 & 0.362 & 0.537 & 0.497 & 0.867 & 0.833 \\
			GST SSIM & 0.386 & 0.635 & 0.461 & 0.715 & 0.516 & 0.722 & 0.698 & 0.835 & 0.743 & 0.833 & 0.797 & 0.817 \\ \hline
			MS-SSIM \cite{wang2003multiscale} & 0.305 & 0.260 & 0.296 & 0.238 & 0.416 & 0.338 & 0.439 & 0.393 & 0.578 & 0.561 & 0.706 & 0.696 \\ 
			GST MS-SSIM & 0.505 & 0.682 & 0.495 & 0.769 & 0.579 & 0.796 & 0.704 & 0.844 & 0.775 & 0.841 & 0.832 & 0.838 \\ \hline
			ST-RRED \cite{soundararajan2012video} & 0.305 & 0.275 & 0.296 & 0.206 & 0.612 & 0.613 & 0.584 & 0.513 & 0.650 & 0.604 & 0.755 & 0.696 \\  
			GST ST-RRED & 0.518 & 0.698 & 0.421 & 0.664 & 0.580 & 0.814 & 0.684 & 0.833 & 0.752 & 0.816 & 0.888 & 0.885 \\ \hline
			SpEED \cite{bampis2017speed} & 0.432 & 0.273 & 0.410 & 0.233 & 0.439 & 0.292 & 0.546 & 0.390 & 0.578 & 0.471 & 0.758 & 0.739 \\ 
			GST SpEED & 0.645 & 0.744 & 0.616 & 0.759 & 0.577 & 0.745 & 0.723 & 0.789 & 0.787 & 0.827 & 0.860 & 0.867 \\ \hline
			VMAF \cite{VMAF2016} & 0.250 & 0.368 & 0.362 & 0.471 & 0.630 & 0.680 & 0.734 & 0.793 & \textbf{0.860} & 0.868 & 0.818 & 0.816 \\
			GREED & \textbf{0.727} & \textbf{0.822} & \textbf{0.702} & \textbf{0.843} & \textbf{0.732} & \textbf{0.840} & \textbf{0.818} & \textbf{0.896} & \textbf{0.864} & \textbf{0.891} & \textbf{0.888} & \textbf{0.895} \\ 
			GREED-VMAF & \textbf{0.748} & \textbf{0.805} & \textbf{0.743} & \textbf{0.833} & \textbf{0.773} & \textbf{0.836} & \textbf{0.786} & \textbf{0.880} & \textbf{0.860} & \textbf{0.899} & \textbf{0.881} & \textbf{0.903} \\ 
			\hline
		\end{tabular}
	}
\end{table*}

To lend a more local nature to the entropy estimates, and to also provide numerical stability in regions having low variances, while preferring more interesting regions of greater activity, the entropy estimates $h(\Tilde{B}_{kpt})$ are scaled by their local variances as
\begin{align*}
    \epsilon_{kpt} = [\log(1+\sigma^2(\Tilde{B}_{kpt}))]h(\Tilde{B}_{kpt}),
\end{align*}
where $B_{kpt}$ the denotes band-pass coefficients corresponding to frame $t$, patch $p$ and band-pass filter $k$, and $\sigma^2$ denotes variance. The changes in entropy values can arise due to two lossy operations : temporal subsampling and compression. To achieve effective quality prediction, a quality model must account for a combination of these operations. However, in \cite{madhusudana2021greed} it was observed that entropy changes due to variations in frame rate often dominated those arising from compression, \ie, are biased towards VFR distorions. Thus, simple comparisons between the entropies of reference and distorted sequences is often insufficient to capture the overall quality. To overcome this shortcoming, a "debiasing" stage is implemented, which requires defining an additional sequence "Pseudo Reference" video $PR$ in addition to the reference video $R$ and distorted video $D$ . $PR$ is obtained by temporally subsampling $R$ to have the same frame rate as $D$. For cases where $R$ and $D$ have the same frame rates, then $PR$ is the same as $R$. Using these signals TGREED is defined as

\begin{align}
    \text{TGREED}_{kt} = \frac{1}{P}\sum_{p=1} ^P \Bigg|\Big(1 + |\epsilon_{kpt} ^D - \epsilon_{kpt} ^{PR}|\Big) \frac{\epsilon_{kpt} ^R + 1}{\epsilon_{kpt} ^{PR} + 1} - 1 \Bigg|.
    \label{eqn:GTI}
\end{align}
This expression can be understood by decomposing it into two components - absolute difference term, and the ratio term. The ratio term is dependent only on the frame rate of distorted video $D$, and is invariant to compression effects, since $D$ is not required for its calculation. On the other hand, the difference term mainly accounts for compression artifacts, since $PR$ can be considered as a lossless version of $D$. The entire TGREED index can be considered as a weighted entropic difference.
\vspace{-8pt}
\subsection{Feature Integration}
\label{sec:feat_integration}

The main goal of this work is to show that using existing FR VQA models that perform poorly when VFR distortions are present, can be transformed into more powerful models able to address frame rate distortions. We accomplish this using a fusion framework, combining spatial features from existing VQA methods, and TGREED features which account for temporal artifacts. Current state-of-the-art (SOTA) VQA models, like those studied here exhibit excellent performance when evaluated on legacy databases like LIVE-VQA \cite{seshadrinathan2010study}, CSIQ-VQA \cite{vu2014vis3} etc. These datasets contain reference and distorted videos having the same frame rates and, include common distortions such as compression, scaling, packet loss etc. However, when evaluated on VFR sequences, these models under-perform due to the presence of frame rate artifacts, as evidenced in Table \ref{Table:MOS_comparison}.

Our feature fusion method is motivated by the successes of earlier fusion-based VQA models \cite{VMAF2016,bampis2018spatiotemporal,madhusudana2021pcs}. Combining features exploits the advantages offered by both models into a single framework. We retain the spatial components present in the current models, and replace the temporal components (if any) with TGREED features. We refer to these modified VQA versions using the Generalized Space Time (GST) prefix. TGREED features are computed at two scales $s=4,5$, where the frames are spatially subsampled by $2^s$ times before calculating the entropy values. The above subsampling was observed to have the optimal trade-off in terms of computational complexity and performance. A biorthogonal-2.2 wavelet filter with three levels of wavelet packet decomposition \cite{coifman1992entropy} was used to obtain temporal band-pass coefficients. Ignoring the low pass subband, 7 TGREED features from the respective temporal subbands were calculated at each scale, yielding a total of 14 features from both scales. For each tested VQA model, a single value corresponding to the spatial component was computed, then concatenated with the TGREED features, resulting in a total of 15 features. These features are then fed to a Support Vector Regressor (SVR), and a mapping from feature space to quality scores is learned using ground-truth subjective opinion scores. The entire feature fusion methodology is summarized in Algorithm \ref{alg1}. The code implementations used in this work have been made available online\footnote{\url{https://github.com/pavancm/GST_VQA}}.

\begin{table*}[t]
\caption{Performance comparison of various FR methods on multiple VQA databases. These databases do \textbf{not} contain variable frame rates. The reported numbers are median values over multiple iterations of randomly chosen train-test sets. In each column the first and second best values are boldfaced. We obtained numbers from \cite{madhusudana2021greed,madhusudana2021pcs} for the results of GREED and GREED-VMAF.}
\label{table:MOS_VQA}
\centering
    \begin{tabular}{|c||cc|cc|cc|cc|cc|}
        \hline
        \multirow{2}{*}{} & \multicolumn{2}{|c|}{LIVE-VQA \cite{seshadrinathan2010study}}  & \multicolumn{2}{|c|}{LIVE-mobile \cite{moorthy2012video}} & \multicolumn{2}{|c|}{CSIQ-VQA \cite{vu2014vis3}} & \multicolumn{2}{|c|}{ETRI-LIVE STSVQ \cite{lee2021subjective}} & \multicolumn{2}{|c|}{AVT-VQDB-UHD-1 \cite{rao2019avt}}\\ \cline{2-11}
        ~ & SROCC$\uparrow$ & PLCC$\uparrow$ & SROCC$\uparrow$ & PLCC$\uparrow$ & SROCC$\uparrow$ & PLCC$\uparrow$ & SROCC$\uparrow$ & PLCC$\uparrow$ & SROCC$\uparrow$ & PLCC$\uparrow$ \\ \hline \hline
        SSIM \cite{wang2004image} & 0.802 & 0.791 & 0.798 & 0.870 & 0.705 & 0.724 & 0.512 & 0.549 & 0.777 & 0.693 \\
        GST SSIM & 0.747 & 0.754 & 0.875 & 0.896 & 0.836 & 0.821 & 0.820 & 0.860 & 0.804 & 0.820 \\ \hline
        MS-SSIM \cite{wang2003multiscale} & \textbf{0.830} & \textbf{0.804} & 0.800 & 0.874 & 0.757 & 0.755 & 0.498 & 0.473 & 0.784 & 0.715 \\
        GST MS-SSIM & 0.745 & 0.730 & 0.879 & 0.888 & 0.873 & 0.847 & 0.831 & 0.845 & 0.807 & 0.820\\ \hline
        ST-RRED \cite{soundararajan2012video} & \textbf{0.826} & \textbf{0.828} & 0.882 & 0.935 & 0.813 & 0.793 & 0.514 & 0.562 & 0.837 & 0.778 \\
        GST ST-RRED & 0.793 & 0.794 & 0.880 & 0.925 & \textbf{0.890} & \textbf{0.927} & \textbf{0.873} & \textbf{0.893} & 0.820 & 0.822 \\ \hline
        SpEED \cite{bampis2017speed} & 0.801 & 0.815 & 0.886 & \textbf{0.929} & 0.743 & 0.743 & 0.466 & 0.552 & 0.834 & 0.785 \\ 
        GST SpEED & 0.782 & 0.774 & 0.890 & \textbf{0.927} & 0.793 & 0.776 & 0.842 & 0.860 & 0.821 & 0.834 \\ \hline
        VMAF \cite{VMAF2016}& 0.794 & 0.794 & \textbf{0.897} & 0.902 & 0.618 & 0.623 & 0.660 & 0.651 & \textbf{0.872} & \textbf{0.873} \\ 
        GREED & 0.750 & 0.755 & 0.863 & 0.908 & 0.780 & 0.766 & 0.730 & 0.760 & 0.823 & 0.823 \\
        GREED-VMAF & 0.784 & 0.786 & \textbf{0.904} & 0.923 & \textbf{0.907} & \textbf{0.899} & \textbf{0.860} & \textbf{0.866} & \textbf{0.842} & \textbf{0.853} \\
        \hline
    \end{tabular}
\end{table*}

\section{Experiments and Results}
\label{sec:experiments}

\subsection{Experimental Settings}
\textbf{Compared Methods} : For performance comparison we choose 5 popular VQA models : SSIM \cite{wang2004image}, MS-SSIM \cite{wang2003multiscale}, ST-RRED \cite{soundararajan2012video}, SpEED \cite{bampis2017speed} and VMAF \cite{VMAF2016}. Of these, the first two are IQA (frame based) models, which are computed on every frame, then temporally averaged to obtain final video quality scores. The above models require the reference and distorted sequences to have same frame rate, thus, when comparing videos with differing frame rates, the distorted sequences were temporally upsampled using frame duplication to have the same frame rate as the reference. In addition to the above models, we also included the recently proposed GREED-VMAF model \cite{madhusudana2021pcs}, which combines GREED and VMAF features to obtain better performance. We employ commonly used evaluation metrics for VQA models, such as Spearman's rank order correlation coefficient (SROCC), Kendall's rank order correlation coefficient (KROCC), Pearson's linear correlation coefficient (PLCC), and the root mean squared error (RMSE). 

\vspace{-7pt}
\subsection{Performance Comparison on the LIVE-YT-HFR Dataset}
The first experiment we performed was to analyze the impact of adding TGREED features to VQA models in regards to capturing frame rate artifacts. In the analysis, we used the LIVE-YT-HFR dataset \cite{madhusudana2021liveythfr}, which contains a total of 480 videos spanning 6 different frame rates. For training purposes, the LIVE-YT-HFR dataset was randomly partitioned into 70\%/15\%/15\% sets for training, validation and testing, respectively. We also ensured that there was no overlap of contents between the subsets. To avoid bias towards any specific train-test combination, the experiment was repeated over 200 different random train-test splits, and the median performance across these combinations is reported. The performances of the different models is compared in Table \ref{Table:MOS_comparison}. For ease of comparison, the original VQA models and corresponding GST versions are clustered together. From the Table, it can be observed that inclusion of the TGREED features provides significant boosts in performance of all the compared models.

In order to analyze model correlations on each individual frame rate, the LIVE-YT-HFR dataset was divided into subsets containing videos having a single frame rate with the resulting performances on each subset reported in Table \ref{Table:FPS_comparison}. From the Table it is evident that there was significant enhancement in performance due to the inclusion of TGREED features on each frame rate, especially low frame rates. 

\vspace{-7pt}
\subsection{Performance Comparison on other VQA Databases}
To analyze the generalizability of the proposed design, we conducted performance comparison of the compared models on traditional VQA databases containing a variety of spatial and temporal artifacts, but single, shared frame rates. For evaluation purposes, we used 5 such legacy VQA datasets, LIVE-VQA \cite{seshadrinathan2010study}, LIVE-mobile \cite{moorthy2012video}, CSIQ-VQA \cite{vu2014vis3}, ETRI-LIVE STSVQ \cite{lee2021subjective} and AVT-VQDB-UHD-1 \cite{rao2019avt}. These datasets span a variety of spatial resolutions, contain various types of distortions such as scaling, compression, white noise etc., and employ multiple compression schemes such as MPEG, H264, H265, VP9 etc. Thus an exhaustive evaluation on them should validate the generalizability of a GREED-enhanced objective VQA model. Note that all the videos in the above datasets have reference and corresponding distorted videos of identical frame rates. Although some datasets contain distorted videos at different frame rates, the authors reported that appropriate temporal processing was applied so that the videos viewed by the human subjects had the same frame rate as the reference.

Each dataset was divided into 80\% and 20\% non-overlapping subsets based on content for training and testing, respectively. This procedure was repeated over all possible train-test splits, and the median performance values reported in Table \ref{table:MOS_VQA}. ETRI-LIVE STSVQ and AVT-VQDB-UHD-1 datasets contain many more source videos, thus for these two datasets the train-test splits were repeated 200 times, similar to LIVE-YT-HFR. From the Table it may be observed that including TGREED features leads to better correlations on most of the compared datasets, and a few of the modified versions also achieve top performance on certain datasets. These results highlight the generalization capability of the TGREED index as well as the efficacy of including effective temporal features for objective video quality prediction.

\section{Conclusion}
\label{sec:conclusion}
We proposed a feature fusion method to improve existing VQA models to be able to capture frame rate distortions. The design approach exploits the superior prediction power of the temporal components present in GREED. This flexible framework is quite generic and can be applied to any existing FR VQA models, including IQA models. We performed a comprehensive evaluation on the LIVE-YT-HFR dataset, and showed that fusing TGREED with existing SOTA models is highly beneficial for improving quality predictions on videos afflicted with frame rate distortions. We also evaluated the generalizability of the proposed method on legacy fixed frame rate databases and observed that the TGREED-enhanced models often achieved better performance than original models. Our results indicate that using powerful temporal representations can lead to better VQA models for both VFR and non-VFR scenarios. These VQA models can be beneficial in determining frame rate dynamically in a constrained bandwidth environment.

\ifCLASSOPTIONcaptionsoff
  \newpage
\fi

\bibliographystyle{IEEEtran}
\bibliography{template_4}

\end{document}